
\input harvmac

\Title{HUTP-91/A057}{Physics Focus and Fiscal Forces}


\centerline{Sheldon L. Glashow}
\bigskip\centerline{Lyman Laboratory of Physics}
\centerline{Harvard University}\centerline{Cambridge, MA 02138}

\vskip .3in
Two items are reproduced herein: my `Outlook' talk, an amended version of
which was
presented at the 1991 joint Lepton--Photon and EPS Conference in Geneva,
and an Open Letter addressed to HEPAP. One is addressed primarily to the
European high--energy physics community, the other to the American.
A common theme of these presentations
is a plea for the
rational allocation of the
limited funds society provides
for high--energy physics research. If my `loose cannon'
remarks may seem irresponsible to some of my colleagues, my silence would
be more so.

\Date{11/91} 

\newsec{OUTLOOK}

Just a few months after CERN's triumphant observations of $Z^0$ and
$W^\pm$ bosons, Harvard celebrated the centenary of its Jefferson Physical
Laboratory. Carlo Rubbia, then one of our most illustrious faculty members,
described these discoveries and more: he spoke with gusto of certain
curious events --- {\it monojets}
 --- which could not be explained with the
standard model of particle physics. We have learned a great deal since
then. There are no inexplicable monojets and there has not been one
demonstrable failure of the standard model. QCD and the electroweak theory
reign supreme. They offer a complete, correct and consistent description of
all known elementary--particle phenomena. Rubbia is now the
director--general of a laboratory that seems forever destined to confirm
the predictions of what we must now call the standard {\it theory.} His
punishment is fit as a modern--day myth.\smallskip

Some physicists look upon the current situation as a triumph of human
ingenuity, since many of the most puzzling problems of the past are solved.
The rich and complex spectroscopy of hadrons is `understood'  at last in
terms of the energy levels of a system of two or three  interacting quarks,
much as the spectroscopy of nuclei is understood in terms of constituent
nucleons and that of atoms in terms of electrons. (The quotes are needed
since we cannot as yet perform as  precise calculations with QCD as we
might wish to do.) Similarly, all of the wealth of weak--interaction
phenomena is resolved in terms of the gauge interactions of three
intermediate vector bosons, whose properties have been found to agree with
the predictions of the electroweak theory.\smallskip

Many questions lie beyond the ken of the standard theory. They
remain to be answered and suggest the existence of a more powerful theory.
Why are there three fermion families? Why do particles have
the masses and mixings they do? Why did nature choose the gauge group
$SU(3)\times SU(2)\times U(1)$? Look to history, say the optimists. The
secrets of the outer atom were exposed at energies of a few eV. Those of
the inner atom by $X$ ray studies at energies of a few KeV. The nucleus was
explored at MeV energies and evidence for quarks appeared at some GeV. Each
great breakthrough required a leap in energy by a factor of 1000. Thus, the
next great discovery to be made --- the nature of electroweak symmetry
breaking and the origin of mass --- will be revealed in the multi--TeV
domain: at the next great hadron collider.\smallskip

Other physicists interpret the manifold empirical successes of the
standard theory as a tragedy marking the end of an era of exciting
discovery. Aside from one missing quark and the Higgs boson, nothing is
left to discover at the high--energy frontier.  All that remains is the
measurement of the next decimal place. Although we have heard this refrain
before, the
pessimists also consult the history of our discipline. They point to the
great surprising discoveries of the past century, which until recently
have sprung upon us every few years:\smallskip

{\bf 1890s\quad} $X$ rays, rare gases, radioactivity and the electron.

{\bf 1900s\quad} Planck's $h$,  half--lives, photoelectric
effect, special relativity.

{\bf 1910s\quad} Cosmic rays, nuclei, the Bohr atom and the bending of
starlight.

{\bf 1920s\quad} Hubble shift, quantum theory, spin and the  Pauli principle.

{\bf 1930s\quad} Positrons, muons, neutrons, fission, continuous beta spectra.

{\bf 1940s\quad} Pions, strange particles, the Lamb shift and the  atomic bomb.

{\bf 1950s\quad} Parity violation, pion--nucleon resonance, neutrinos,
$V-A$.

{\bf 1960s\quad} Muon neutrinos, scaling behavior,
CP violation, hadrons aplenty.

{\bf 1970s\quad} $J/\Psi$, neutral currents, tau leptons, charm
and beauty quarks.

{\bf 1980s\quad} SN1987a, voids and walls of the universe.\smallskip

\noindent While new and wonderful things are being found out about the
universe, there hasn't been a big surprise in our discipline since
the establishment of the third family of quarks and leptons in 1977. True,
$W$ and $Z$ were first produced and detected in the 1980s, but their
reality came as no surprise.  Not for well over a century has there been
such an insipid interregnum. So be it. While optimists glory in the
success of the standard theory and pessimists bemoan their fate, there
remains much to be done:\smallskip

$\bullet$ Let us not be hasty in assuming its absolute validity: the
standard theory must be {\it tested\/} as best we can. With a million $Z^0$
events in hand, the theory works just fine. Will agreement persist when we
have  ten times that number?  Will the deduced value of the muon's $g-2$
survive the next experimental onslaught? Is the Kobayashi--Maskawa matrix
truly unitary? And so on. \smallskip

$\bullet$ We must learn to {\it exploit\/} the standard model better.
Theorists and experimenters must work together to devise more and better
ways to extract the consequences of a recalcitrant  theory and confront
them with experiment.\smallskip

$\bullet$ We must {\it hope\/} for more surprises and search for them.
Nature is not the enemy; complacency is. No physicist should be so arrogant
as to believe that her bag of tricks is exhausted, lest the great desert of
particle physics  become a self--fulfilling prophecy. \smallskip

$\bullet$ We must {\it worry\/} the weak points of the standard theory. CP
violation is surely among them. Another generation of experiments is needed
to pin down the fundamental CP--violation parameters in $K^0$ decay. The
search for a neutron electric dipole moment must go on. Most importantly,
the world needs one (and only one!) $B$ factory powerful enough to explore
the question of CP violation in $B$ decay.

$\bullet$ We must {\it extend\/} the burgeoning non--accelerator frontier
and not forget that much of what we know about elementary particles
once came, and will come again,
from disciplines far removed from accelerator--based high--energy
physics:\smallskip

\item{\bf (a)} The existence of an alleged 17 KeV neutrino is flatly
inconsistent with the standard theory. Does it exist or does it not?
Discussions at this meeting were inconclusive. Many experiments, some of
them solid, see evidence for it. Many experiments, some of them solid,
claim to exclude it. Here is an exciting dilemma! \item{\bf (b)} Does
resolution of the solar--neutrino problem demand the existence of mixed and
massive neutrinos? These effects lie beyond the minimal 17--parameter
standard theory. However, the case for MSW oscillations as a cure to the
problem is far from iron clad. Morrison, for example, is led to conclude
that {\it there is no solar neutrino problem.} Decisive results from the
gallium experiments are awaited with fervor, as are those from future
experiments now being planned. \item{\bf (c)} Astronomers have come to the
dreadful conclusion that they cannot detect anything beyond the
gravitational influences of the dominant form of matter in the universe.
The dark matter may be non--baryonic and unconventional. If so, its
identification and investigation lies within the realm of high--energy
physics.\smallskip

The near--term future of elementary--particle physics, largely if not
exclusively, depends on those large accelerators that are now or soon
operating: the fixed target facilities at Fermilab, Brookhaven's kaon beam,
the Tevatron collider, LEP, HERA and the mini $B$ factory at Cornell.  I
have not the gall to write a guide for the experimenter. Our discipline
is open--ended, in the sense that the new and profitable directions are
almost certainly not those that we anticipate. Surprises may lie in wait
for experimenters at all of these facilities.\smallskip

For the long term,
our community is almost unanimous in its belief that the next great
accelerator must be a hadron collider. SSC is approved and partially
funded while an LHC proposal
is soon to be put to the CERN council. At the risk of
alienating many friends and colleagues, let me consider this
issue.\footnote{$^\dagger$}{The following remarks on the issue of one too
many supercolliders were not a part of my oral
presentation at the conference.}\smallskip

Giant accelerators, their enormous detectors, and their continuing
operations are enormously expensive enterprises. {\it We all agree that the
world needs one great hadron collider, but two such machines makes no sense
at all. The world's governments, if they knew what we know, should not fund
the construction and instrumentation of both machines. There are too many
other interesting and important things to do!}  Although  well--intentioned
arguments have been presented for each accelerator from each side of the
Atlantic, the course that is set seems bound for disaster. \smallskip

Should we prove ingenious (and  irresponsible) enough  to convince our
governments to begin building  the {\it two\/} comparable behemoths, these
endeavors are likely to starve ongoing research programs in all of physics,
and especially in high--energy physics. Money is more--or--less conserved:
CERN {\it needs\/} extra support (from member or non--member) nations to
build LHC expeditiously, and SSC may not be built without substantial
external support. The majority of our colleagues --- those who do physics
in real time and choose not to live at the  very fringe of the high--energy
frontier --- will need and deserve funds that are almost certainly  to be
pre\"empted by our preoccupation with bigness. \smallskip

If and when the twin supercolliders begin to do real physics research ---
which may not happen until the next millenium ---  there may no longer
exist enough of a community of experienced and dedicated high--energy
physicists to use them.  It is not for reasons of national pride that I
believe that SSC is the preferred machine if we have the vision, the
scientific integrity, and especially, the fiscal responsibility to build
only one.\smallskip

$\bullet$ There are {\it technical obstacles\/} in the path of the
construction and exploitation of LHC. It requires  novel two--in--one
magnets operating  at nearly ten tesla. Can reliable and
adequate magnets can be industrially produced at reasonable cost? The
physics reach of the LHC is extended by exchanging a compromised collision
energy for an intense luminosity, but can appropriate detectors be designed
and built to make use of such luminosities?

$\bullet$ There are {\it operational difficulties\/} associated with the
construction of LHC. LEP has made a glorious start in
its ambitious
and exciting program. It has both a higher--luminosity and a higher--energy
phase to look forward to. It would be tragic for physics if LHC deployment
would hamper, delay or constrain these essential projects.

$\bullet$  There is a {\it serious scientific risk\/} associated with the
exploitation of LHC. Its collision energy is severely
constrained by its need to fit within the LEP tunnel. What if
the new physics is accessible at 40 TeV but not at 20 TeV?
\smallskip

My dream is of a `new world order' of cooperation rather than
competition for which high--energy physics can lead the way. Let SSC evolve
into a truly international project: in funding, instrumentation and
administration. CERN is an enormously successful and enviable model. Let
SSC fly the flags of all interested and committed nations and be
administered by a representative council  much as CERN is. Let it be a
machine {\it in\/} America but {\it of\/} the world. My dream is certainly
naive, unrealistic and all but impossible to realize. Until recently, the
DOE has insisted that SSC be an essentially American enterprise. CERN needs
a new initiative to ensure its continued vitality. Nonetheless,  I cherish
the hope --- for the sake of our discipline --- that my dream may come
true.\smallskip

Much has been said at this conference about grand unification. Strong, weak
and electromagnetic forces are each mediated by gauge bosons. The
hypothesis that there is a simple underlying gauge group is irresistably
attractive. In minimal $SU(5)$, the curious charges of quarks and leptons
are forced upon us and each family corresponds to an anomaly--free
representation. Neutrinos are automatically massless and neutral. The
observed disparity in strength between strong and electric forces sets the
symmetry--breaking scale, makes protons practically stable, and evaluates
the weak mixing angle. The trouble is in the details: protons are not
stable enough and $\sin^2\theta$ comes out a bit too small.\smallskip

There are many ways in which minimal $SU(5)$ may be modified so
as to patch up the problems, fit the data, and save the notion of
grand unification.
 {\it The First Fix: } A
decade ago, several theorists pointed out that supersymmetrized $SU(5)$
could do the trick
\ref\rmsss{S. Dimopoulos and H. Georgi, Nucl. Phys.
B193(1981)150. S. Dimopoulos, S. Raby and F. Wilczek, Phys. Rev.
D24(1981)1681. L.E. Ibanez and G.G. Ross, Phys. Lett. B105(1981)439.}.
{\it The
Second Fix: } Both empirical problems could be dealt with by adding to the
standard theory one or more split $SU(5)$ fermion
multiplets \ref\rsplit{{\it E.g.,} P. Frampton and S.L. Glashow, Phys.
Lett. B135(1983)340.}.  {\it The Third Fix: } Perhaps there is an
intermediate stage of symmetry breaking involving a semisimple gauge group
lying between the unifying group and the group of the standard theory. An
intersesting example of such a hierarchy is:
$$O(10)\rightarrow SU(3)\times SU(2)\times SU(2)\rightarrow SU(3)\times
SU(2)\times U(1).$$
The intermediate mass scale can be chosen to fit the low--energy values of
$\alpha,\,\alpha_s$ and $\cos\theta$ \ref\rgh{H. Georgi and D.V.
Nanopoulos, Nucl. Phys. B159(1979)16. H. Georgi ans S. Dawson, Nucl.
Phys. B179(1981)477.}.
For this case, the unification scale
is comparable to the Planck energy (neat, but no observable proton
decay), and left--right symmetry is restored at roughly a million TeV
(useful to generate sensible neutrino masses, but not for new phenomena at
LHC or SSC).

Now that LEP data has
provided us with an accurate determination of $\alpha_s$, many
workers have refocussed on the rescue of grand unification and the
intriguing possibility
that lots of new particles lie almost within reach
\ref\rall{{\it E.g.,}
U. Amaldi, W. de Boer, H. F\"urstenau, Phys. Lett. B260(1991)447.
A. Giveon,
L. J. Hall and U. Sarid, Berkeley preprint. F. Anselmo, L. Cifarelli, A.
Petermann and A. Zichichi, CERN preprint. P. Langacker and M. Luo, Univ. of
Penn. preprint.
{\it See also:} S. Dimopoulos, S. Raby and F. Wilczek,
Physics Today (Oct. 1991)25.}. The results  of their
analyses are remarkable (and have received lots of attention in the
semi--popular press).
\smallskip

Although the failure of minimal $SU(5)$ is now established beyond any
possible doubt, its supersymmetrized version is still very much alive.
Furthermore, the super--partners of known particles {\it may\/} lie
$\sim\!1$ TeV and be accessible to the next great hadron collider; and
proton decay {\it may\/} be detectable at super--Kamioka. A  careful and
current analysis  of the data
\ref\rfram{U. Amaldi, W. de Boer, P.
Frampton, H. F\"urstenau, and J. Liu, private communication.} is dramatically
summarized in figure (1).\footnote{$^\dagger$}{The figures are omitted from
this preprint. All three fixes work flawlessly.}
However, the second fix (which does not please supersymmetry fans, but
involves many fewer hypothetical and perhaps accessible particles) is also
in the running, as figure (2) demonstrates. The first and second
fixes {\it
suggest\/} --- the input data are not precise enough as yet to use a
stronger verb --- the possible existence of new physics at soon--to--be
accessible energies. However,   the third fix also preserves grand
unification with an intermediate mass scale lying well beyond  experimental
reach.  And, there may be other fixes as well. We shall learn more about
the various possibilities as data improve, and we cannot
emphasize too strongly the importance of those `pedestrian' experiments, at
LEP or HERA, that serve to measure the value of the strong coupling
constant. \medskip

In summary, let me quote a countryman, {\it `The game ain't over 'til it's
over.} Particle physics is very much alive, both at and away from the
highest energies, both at and away from accelerators. Our colleagues
who lean toward astrophysics and cosmology have given us a solar neutrino
headache and various costly medicines to cure it, gravitational lenses, an
indirect confirmation of gravitational radiation and a hoped--for (but
unfunded) search for such waves,  supernova neutrinos, a weird large--scale
structure to the universe coupled with a perfectly smooth era of
recombination, inexplicable burster phenomena, the possibility of
unconventional dark matter, and many more surprises and dilemmas to come.
Our low--energy colleagues give us monopole constraints, axion searches, an
alleged 17 KeV neutrino, an exorcism of the fifth force, a
precision test of the electroweak theory, and again, more to come.
\smallskip

At the accelerator frontier, there are many important experiments to do and
surprises to uncover. Indeed, there is far more to do than we can easily
afford, especially with the burden of building and instrumenting
supercolliders in two continents. The Fermilab collider must be upgraded so
that it may do its best to find the top, which will be a primary background
to experiments at larger hadron colliders. Fixed--target experiments are
needed to measure $\epsilon^\prime$ and to search for rare decays and
neutrino oscillations. HERA may find leptoquarks, but it will certainly
lead to precision tests of QCD. LEP, in its present and future avatars,
 may yield wonderfully inexplicable
events, but it offers the best tests of the electroweak theory and the best
constraints on its variations. A $B$ factory is essential if we are to
clarify our understanding of quark mixing and CP violation, while smaller
factories which are dedicated to
physics at 1 GeV and at 3 GeV have important r\^oles
to play as well. Finally, and in my view most importantly, we must learn
how to build a powerful linear  electron--positron collider with which  to
study the TeV domain in a relatively clean environment. \smallskip

Many of our best theorists are all wrapped up in strings and conformal
field theory. They believe that a new mathematical framework is needed that
lies beyond tried and trusted quantum field theory. They are almost
certainly right in their belief: the most puzzling {\it `why questions'}
cannot even be posed in the standard theory or any of its conventional
modifications. However, their early optimism is  gone: most string
theorists no longer believe that a theory of everything lurks around the
corner. They should not be discouraged. The new theory will come in its
time, and today's theorists may lead us to it kicking and screaming.   In
the meanwhile, {\it particle physics remains, as it traditionally has been,
primarily an exciting and rewarding experimental science.}

\newsec{AN OPEN LETTER TO HEPAP}

Esteemed colleagues: \
Some years ago, I became a member of HEPAP. For all I know, I may yet be. I
attended the first two meetings during my tenure, and no others. The
reasons include time pressure and my own irresponsibility, but the primary
cause for my delinquency is a strong impression that HEPAP (unlike the CERN
Science Policy Committee, in which I served  honorably for six years) does
not adequately address the needs of the community. Due to the open nature
of its meetings, a constituency that reflects the conflicting
self--interests of our several national laboratories,
the obligation to follow a firmly orchestrated agenda, the
absence of prior discussions and the lack of executive sessions, difficult
issues are rarely faced and sad truths  are withheld
or sloughed over. I am led to write
this letter --- rather than play--act at HEPAP --- because of my love for
and concern about the discipline of accelerator--based high--energy
experimental physics in the United States.

I apologize for this hurried and distraught communication, but I am
alarmed by the precipitous decline of a discipline that Americans created
and, until recently, dominated. There is not reason
nor sense for continuing American domination of the field, but we should
at least remain as major players. We spend a lot of money on HE physics, but
seem to get too little physics done per dollar spent. If we are to
argue, as I think we should, for more funding --- or even for continuing
funding at present levels --- we must demonstrate more of a sense of fiscal
responsibility than we have in the past.  At the price of alienating many
colleagues and friends, let me say the unsayable:\smallskip

$\bullet$ The SLD detector at SLAC has been a sink--hole for funds that
could have been spent productively: it is a second detector for a machine
that will probably never do useful physics. (CERN has analyzed a million,
going on ten million, $Z^0$ events. SLC has got a few hundred, and won't do
all that better, polarization or no. Face it chums: it's a brilliant
demonstration of a new accelerator technology, but will never be much of a
research tool.) Years ago, when the SLD project was initiated, its
proponents may have had sound arguments for it. Very soon, however, it
became clear to much of the community that the device was pointless.
However, once the great bureaucratic ship sets its sails, no agent on
Earth can correct its course. Physicists lucky enough to be
on board cling blindly to to their challenging tasks no matter that the
cruise is to nowhere. For all I know, money is being spent on this useless
toy even as I write.\smallskip

$\bullet$ When the Soudan proton decay detector was first proposed, it
seemed a good idea. If the larger IMB and Kamioka could detect proton
decay, a smaller but more highly instrumented device would have been  an
invaluable facility. As it turned out many years ago, there was no such
signal. It became clear that the Soudan initiative was pointless, as far as
its original primary goal is concerned. No matter. The ship plows on.
\smallskip

$\bullet$ Although PEP was in many ways a better instrument than PETRA, its
timing was a disaster. Aside from a few physics gems, it came on line far
too late to succeed as a forefront accelerator.  As my father taught me, if
a job's worth doing, it's worth doing right... {\it and when it needs
doing!}\smallskip

$\bullet$ Years ago, it seemed a good idea to provide the Fermilab collider
with a better and bigger (and very expensive) new detector. But a second
detector is useful only when there is the potential to exploit the
accelerator at which it is to be used. However, we have hardly begun to
exploit the CDF detector. With collider runs as far apart as they are, who
needs another detector? It's just one more uncontrollable spigot or useless
ship. All that money could have been used to shore up the decaying
university--based infrastructure upon which all of physics research depends.
\smallskip

Enough spilt milk and rotten eggs! High--energy physics in America is
threatened by dilution (too many sacred laboratories), by dispersion (too
many directions, with little foresight and no corrective capacity) and by
the reluctance of our community and its sponsors to perform difficult but
necessary triage. In the future, we must focus more tightly and painfully,
or abandon ship.

For sound scientific and technical reasons,
we all agree that a proton super--collider is our first
priority. However, SSC will not come on line for physics for a
decade. If things continue as they are, {\it there aren't going to be any
American experimenters around who are
ready, willing  and able to use it.} All other
priorities are irrelevant and immaterial unless we properly attend
to numero uno!

As I talk to my experimentalist friends, I find that almost all groups --
from the brilliant to the merely sound --- are being squeezed out of the
business of high--energy physics. Hardly any American  graduate
students care to board our sinking ships. Yet, if we are to build
SSC, we must keep HE physics, especially hadron--collider physics,
healthy in the meantime  --- and there's  lots of good  physics remaining
to do at the World's Highest Energy but Rarely Running Accelerator.
 The training ground for a large
hadron collider is a large hadron collider, and we've got the only one.
We've got to `find the top' because it's likely to be the background
to  really interesting SSC events.
Can anyone doubt that part and parcel of the SSC initiative must be the
maximum exploitation of the Fermilab collider. This means
that {\it we must have more
collider runs sooner and a Main Injector too!}

One last word about the SSC, whose balance sheet I've never quite
understood. It seems that we are {\it depending\/} on  massive foreign
support for the construction of both the machine and its  enormous
detectors. We don't have committments for such support, and we may never do
if Europe commits to the LHC and Japan joins their action. Furthermore, it
is international lunacy to build and instrument {\it both} behemoths.
Arguments based on physics goals and technical obstacles convince me that
SSC is the better machine to build: LHC is just too small, too hard to
build and too hard to instrument. Is the federal government  finally
reaching the sensible conclusion that SSC must be truly internationalized,
not only in funding but in management, so that it  becomes a facility in
America but of and for  the World?  The search for the secrets of matter
and the universe is both glorious and costly. Surely, we must
cooperate rather than compete --- it's the way to go, and perhaps the only
way.

\bigbreak\bigskip\bigskip\centerline{{\bf Acknowledgements}}\nobreak
We gratefully acknowledge communications with P. Frampton and U. Amaldi.
This work was supported in part by
NSF grant PHY-87-14654 and by TNRLC grant RGFY9106.

\listrefs
\bye